\documentclass[useAMS,usenatbib,fleqn,usegraphicx]{mn2e}
\usepackage{amsmath,amssymb}
\usepackage{natbib}
\bibpunct{(}{)}{;}{a}{,}{,}

\title[IC4499]
{Radial Velocity and Metallicity of the Globular Cluster IC4499 Obtained
  with AAOmega
\thanks{Based on observations made with the Anglo-Australian
  Telescope operated at Siding Spring Observatory by the Anglo-Australian Observatory.}}
\author[Hankey \& Cole]
  {Warren J. Hankey$^1$ and Andrew A. Cole$^1$ \\
  $^1$School of Mathematics \& Physics, University of Tasmania,
  Private Bag 37, Hobart, TAS 7001, Australia; warren.hankey@utas.edu.au}

\begin{document}
\maketitle
\begin{abstract}

We present radial velocity and metallicity measurements for the far-southern
Galactic globular cluster IC~4499.  We selected several hundred target red
giant stars in and around the cluster from the 2MASS point source catalog,
and obtained spectra at the near-infrared calcium triplet using the AAOmega
spectrograph.  Observations of giants in globular clusters M4, M22, and M68 were
taken to provide radial velocity and metallicity comparison objects.  Based on 
velocity data we conclude that 43 of our targets are cluster members, by far
the largest sample of IC~4499 giants spectroscopically studied.  We determine
the mean heliocentric radial velocity of the cluster to be 31.5 $\pm$0.4 km/s,
and find the most likely central velocity dispersion to be 2.5 $\pm$0.5 km/s.
This leads to a dynamical mass estimate for the cluster of 93 $\pm$37$\times10^{3}$
M$_{\odot}$.  We are sensitive to cluster rotation down to an amplitude of 
$\approx$1~km/s, but no evidence for cluster rotation is seen.   The cluster
metallicity is found to be [Fe/H] = $-$1.52 $\pm$0.12 on the Carretta-Gratton scale;
this is in agreement with some earlier estimates but carries significantly higher
precision.  The radial velocity of the cluster, previously highly uncertain, is
consistent with membership in the Monoceros tidal stream as proposed by Pe\~{n}arrubia
and co-workers, but also with a halo origin.  The horizontal branch morphology
of the cluster is slightly redder than average for its metallicity, but it is likely
not unusually young compared to other clusters of the halo.  The new constraints
on the cluster kinematics and metallicity may give insight into its extremely high
specific frequency of RR~Lyrae stars.
\end{abstract}


\section{Introduction}

IC~4499 is a sparsely populated globular cluster in a crowded galactic
field near the south celestial pole. It was discovered in 1900 by D. Stewart 
\citep{pic08} and has been comparatively understudied,
probably as its extreme southern declination presents an observational challenge 
to mid-latitude observers.  Several photometric studies of IC~4499 have been 
undertaken to study the horizontal branch (HB) morphology, produce colour-magnitude
diagrams (CMDs) and make distance and metallicity estimates, but no detailed 
spectroscopic metallicity or radial velocity data have been published to date.

The globular cluster catalogue of \citet{harris96} gives a distance of 18.9~kpc,
which puts it 15.7~kpc from the Galactic centre and 6.6~kpc below the plane of the
galaxy, making it an outer halo cluster.  From the vantage point of Earth, it is 
seen through the outer parts of the Galactic bulge ($\ell$ = 307.35$^{\circ}$,
$b$ = $-$20.47$^{\circ}$), resulting in a relatively high reddening.  This reddening
has been estimated as high as E(B$-$V) $\approx$0.35 \citep{fou74}, but more recent
work suggests smaller values of 0.15--0.25
\citep{saraj93,storm04,walker96,ferraro95}.  The uncertainty in reddening has 
likely propagated through into differences in conclusions about the metallicity, 
distance, and age of IC~4499.

Low-resolution spectroscopic radial velocities
and metallicities have been obtained for three RR~Lyrae stars only \citep{perk}.
They obtain a metallicity of [Fe/H] = $-$1.33 $\pm0.3$ from the strength of the
singly-ionised Ca~II~K line using the $\Delta$S method \citep{smith84}.
On the scale of \citet[ZW84]{zinn} this becomes [Fe/H] = $-$1.5 $\pm0.3$.
\citet{fus95} noted a discrepancy between photometric metallicity
estimates, which tend to be around [Fe/H] $\approx$ $-$1.75 \citep{ferraro95}
and the generally higher spectroscopic estimates.
More recent unpublished work by R.
Cannon (1992) is quoted by \citet{saraj93} and \citet{walker96} as yielding [Fe/H] =
$-$1.65 on the ZW84 scale, based on the near-infrared Ca~II triplet lines of four
red giants.
\citet{perk} also published radial velocities for their three RR~Lyrae stars,
 -60, +10 and -101 km/s, all with an error of $\pm$50 km/s. 

IC~4499 is noteworthy in having an extremely high specific frequency of RR~Lyrae
variables; its value of 
S$_{RR}$ = 113.4\footnote{S$_{RR} \equiv$ N$_{RR}$10$^{0.4(7.5+Mv)}$ for a cluster 
of absolute magnitude M$_V$ with $N_{RR}$ variables.}
is second only to the smaller Fornax~1 globular cluster
\citep{mac03} and the tiny outer halo cluster 
Pal~13 \citep{harris96}.  About 100  RR~Lyrae stars have been identified and represent
$\approx$68\% of the the total HB population \citep{saraj93}.  Most of the RR~Lyr
have P $\leq$0.6~d, making it an Oosterhoff Type~I (OoI) cluster \citep{clement,walker96}.
Metallicity may be an important factor in determining the Oosterhoff classification
of a cluster, as most OoI clusters tend to be more metal-rich than [Fe/H] = $-$1.8
on the ZW84 scale, while Oosterhoff Type~II clusters more metal-poor \citep{sand93}. 
It is thought
that shorter-period RR~Lyrae stars have not evolved off of the zero-age horizontal branch
(ZAHB), while the longer-period variables are evolving through the RR~Lyr instability
strip on the way to the asymptotic giant branch.  The measurement of accurate cluster
parameters therefore has the potential to shed light on the evolutionary pathways of
horizontal branch stars \citep[e.g.,][]{clem00,pritzl00}.

IC~4499 has been proposed as a ``young'' globular cluster with an age 2-4 Gyr younger
than 
clusters with similar metallicity \citep{ferraro95}, where age is established by
differential magnitude and colour comparisons with the
main-sequence turnoff (TO).  The method compares magnitude difference between
the HB and TO, and the colour difference between the
red giant branch (RGB)
and TO.  In clusters of similar metallicity, the magnitude difference increases
and the colour difference decreases with increasing age.  \citet{ferraro95} adopt
a value of [Fe/H] = $-$1.8 on the ZW84 scale in their work, and find that IC~4499
is essentially coeval with Arp~2 and NGC~5897.  However, this matter is not settled,
as the similarly-derived compilation of 55 globular cluster ages by \cite{sal02}
finds an age of 12.1 $\pm$1.4~Gyr for IC~4499, not significantly younger than the average of metal-poor clusters. While the latter study assumed the
cluster was 0.3~dex more metal-rich than \citet{ferraro95} did, they arrived at a 
similar conclusion about the cluster coevality with Arp~2 and NGC~5897.  A careful
consideration of the cluster metallicity must be made in order to help resolve this
discrepancy.

\citet{fus95} noted that IC~4499 lies near a great circle around the Galaxy that
passes through other possibly ``young'' globulars, including Pal~12 and Rup~106.
This suggestion is a forerunner of the modern studies of halo substructure based
on searching for tidal streams and RGB overdensities.
In the past decade there has been a rapidly growing awareness of substructures
in the Galactic halo \citep[e.g.,][]{mor00,yan00,viv01,new02}.  Apart from
the tidal stream of the disrupting Sagittarius dwarf spheroidal, one of the strongest
structures detected in photometric surveys is the Galactic Anticentre Stellar Structure,
which is also known as the Monoceros tidal stream or ring \citep[e.g.,][]{new02,iba03}.
The Monoceros stream may be associated with the tidal disruption of a dwarf galaxy close
to the plane of the Milky Way, possibly the Canis Major dwarf
\citep[e.g.,][]{hel03,mar05}; it is also possible  that
the Monoceros stream is a dynamical structure intrinsic to the thick disk of the Milky Way
\citep[e.g.,][]{pia08,you08}.
Several Milky Way star clusters have been suggested as members of the Monoceros stream 
\citep[][and numerous references therein]{mar04,frinch,mon05,pia08,war09},
and this could have strong impacts on studies of the statistics of the
Milky Way globular cluster population if a number of clusters are found to have 
extragalactic origins.

No observational study of the Monoceros stream covers the neighbourhood of IC~4499.
However, a set of numerical models of the stream, developed under the hypothesis that it is
the remnant of a disrupting dwarf galaxy, have been proposed by \citet{mon05}. In one
of their best models, stellar debris stripped from a progenitor dwarf at 
$\ell$ = 245$^{\circ}$, $b$ = $-$18$^{\circ}$ encircles the Milky Way within $\pm$30$^{\circ}$
of the Galactic plane, crossing the location of IC~4499 after nearly a complete wrap.
\citet{mon05} suggested that IC~4499, along with several other clusters, could be 
candidate members of the stream on the basis of their position and the predicted
radial velocities in their models.  The radial velocity of IC~4499 has not yet been
determined accurately enough to check for consistency with this type of model.

Following the methodology of \citet{war09},
we have undertaken a spectroscopic study of IC~4499's red giants in order to obtain
radial velocity and metallicity measurements for a large sample of cluster members; 
the aim is to shed light on questions of its relative age and possible membership in
a stellar stream.  We employ the relationship between CaII triplet
line strengths and [Fe/H] to obtain metallicity estimates for individual giants.
The near-infrared CaII triplet, resulting from absorption by the 
3$^2$D--4$^2$P transition, is a strong feature of
late-type giant stars \citep{arm1988}.  The equivalent width of the lines increases
monotonically with metallicity, regardless of age, for stars older than 1~Gyr
\citep[e.g.,][]{caIIsyn}.

Spectroscopy of the near-infrared
calcium triplet in spectral type K giants has become an accepted tool
for assessing the metallicity of stellar populations \citep{arm91}, being 
calibrated against Galactic globular clusters \citep{rutl97b}.
Originally used in studies of old, simple stellar populations,
the technique has been shown to apply to non-globular cluster stars,
including open clusters and composite populations
\citep[e.g.][and references therein]{cole2004,gro06,bat08}.

The line strength has a strong dependence on surface gravity and a milder
temperature dependence \citep{arm1988,caIIsyn}, which is
removed using the empirical relationship between gravity, temperature,
and luminosity for red giant stars.  \citep{rutl97b} showed that using 
the stellar apparent magnitude with respect to the cluster horizontal branch
is a robust approach to this procedure.  Because of the availability of 
JHK$_S$ photometry and astrometry in the 2MASS catalog, we adopt the 
K-band as our reference magnitude, following \citet{war09}.  Their
relationship between K$-$K$_{HB}$, [Fe/H], and Ca~II equivalent width
is confirmed by our observations of  
IC~4499 and three other clusters.  

We discuss our approach, observations, reductions, and analysis in the next section.
Because using K magnitudes to correct for surface gravity is relatively novel
compared to V or I, we rederive the relation between K$_s$ magnitude above the 
horizontal branch and CaII line strength.  Using three well-studied clusters as a calibration
sample, we present new abundance and radial velocity measurements for IC~4499 in \S 3.
We examine our data for signs of cluster rotation, and rule out rotation velocities in IC~4499 
of 1 km/s or more.
In \S 4 we discuss the implications of our results, including the contention that 
IC~4499 is younger than the bulk of halo globulars \citep{ferraro95,fus95}, and the possibility
that IC~4499 belongs to the Monoceros stellar stream \citep{mon05}.

\section{Methodology}
\subsection{Observations}
Observations were carried out on 28 May 2008 at the 3.9m Anglo-Australian 
Telescope at Siding Spring Observatory. The seeing was 1.4$^{\prime\prime}$.
The AAOmega fibre-fed multi-object spectrograph (MOS)
allows for up to 392 simultaneous spectra to be obtained, across 
a two degree field of view \citep{sharp06}. The 1700D grating was used, which gives
a spectral resolution of about 10,000 [$\lambda/\Delta\lambda$], varying slightly across the field.
This grating spectrum is in the the near-infrared with usable 
 $\lambda\lambda$ 8450 to 8700 \AA, which includes the ionised calcium 
triplet lines at 8662, 8542 and 8498 \AA.  This feature is among the 
strongest lines in K-type giants \citep{arm1986}, which is the dominant spectral
type for red giants at low-metallicity.

\begin{figure}
\includegraphics[width=84mm]{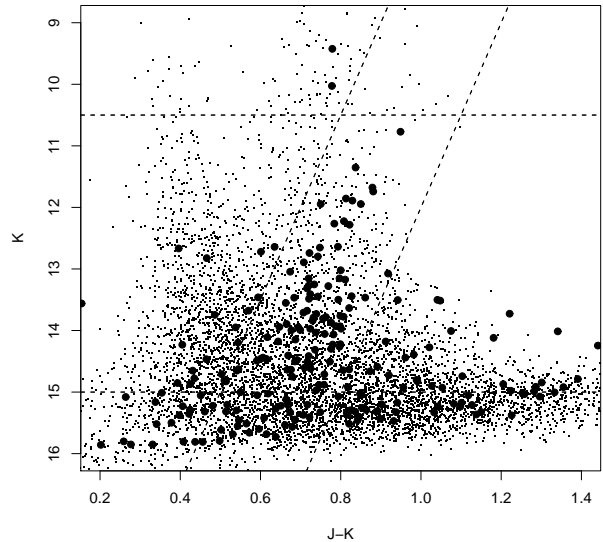}
\caption{Selection of RGB stars from 2MASS PSC within 1$^{\circ}$ of the centre of
IC~4499. Objects within 5$^{\prime}$ of the cluster centre are plotted with large symbols to
highlight the cluster RGB relative to the field.  Our spectroscopic sample is selected
from candidates within the parallelogram containing the cluster RGB.}
\label{fig-cmd}
\end{figure}

Targets were chosen from the 2MASS point source catalogue (PSC),
which has a positional accuracy of about 100 mas \citep{twomass};
this accuracy is crucial to the success of the observations because of the
necessity to accurately place the 2.0$^{\prime \prime}$ fibres on the targets.
The selection was based on 2MASS J and K photometry with K between 10.5 and 15.0.  Because of the slope of the RGB, the color selection had a slope as well, although quite steep.  The red limit was set by $K > 27.0 - 15(J-K)$, and the blue limit was set by $K < 22.5 - 15(J-K)$. The selection region is shown on the CMD in Figure~\ref{fig-cmd}. The highest MOS fibre allocation priority was given to stars within the cluster half-light radius and in the upper 2.25 mag of the RGB.

\begin{figure}
\includegraphics[width=84mm]{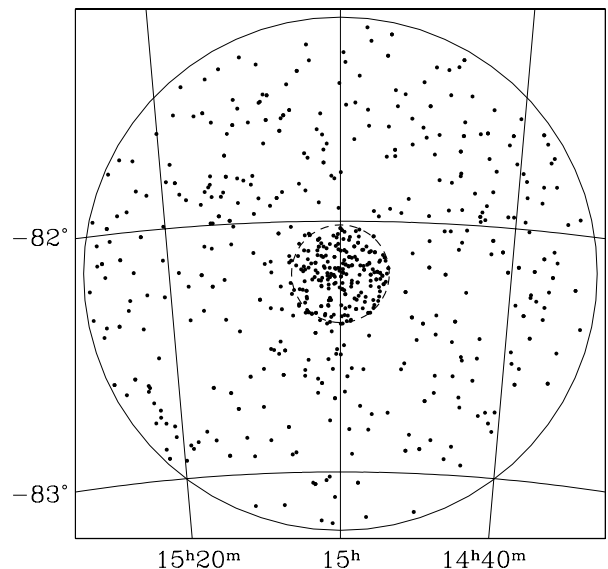}
\caption{Observed targets in a $2^{\circ}$ field around IC~4499. The tidal radius
is shown by the dashed line; the fibre allocation was strongly weighted to select
targets within this radius.}
\label{fig-map}
\end{figure}

The half-light and tidal radii of IC~4499 are 1.5$^{\prime}$ and
12.35$^{\prime}$, respectively \citep{harris96}.  Fibres were preferentially allocated to the
centre of the 2$^{\circ}$ field.  Once the cluster centre was sampled as densely
as possible with fibres, the spare fibres were allocated to stars outside the cluster
centre in the same colour and magnitude range.  This should allow for a very precise
characterisation of the radial velocity distribution of the field stars, to assist
with membership decisions.  Because of the density of the cluster, not all stars could
be observed in a single setup.  We observed two different fibre configurations with the
same central position in order to maximise the yield of members.
In total 569 individual stars were observed with signal to noise $\geq 15$ per pixel in a two degree field around IC~4499; the targets are mapped in Figure~\ref{fig-map}.
The fields were integrated over several exposures to mitigate systematic
errors and cosmic ray contamination.  The total exposure time for each IC~4499
setup was 3600 sec.  

Three well-studied clusters were chosen as comparison objects.  These were picked to 
yield stars of similar spectral type and metallicity to use as radial
velocity templates in our cross-correlation, and to confirm that we could reproduce the
relationship between K magnitude and CaT equivalent width across a range of metallicities.
We used M68 (NGC~4590), M4 (NGC~6121), and M22 (NGC~6656) as our comparison objects;
their positions, relevant properties, and observing details are listed in Table~\ref{tab-log}.
For the calibration clusters, RGB stars were chosen from 2MASS J and K photometry in the regions of the 
selected clusters. (J$-$K, J) CMDs were created for
square-degree areas centred on each cluster, and targets were selected from
the cluster RGB locus down to and including the HB. 
We tried to sample as wide a range of magnitude
as possible in each cluster in order to accurately model the influence of surface
gravity on the CaT equivalent widths.  In general, there are relatively few
bright RGB stars, and the brightest, coolest stars are often contaminated
by titanium oxide bands in the spectral region of interest, so sampling the bright
end of the RGB while respecting the restrictions on minimum fibre spacing was a
challenge.  In most cases, the cluster RGB sequences are not
clearly distinct from the surrounding field, and the samples were cleaned according
to radial velocity and position relative to the cluster centre.  The individual spectra
of each target were coadded after extraction and dispersion correction.

Calibration exposures including arc lamp and screen flats were taken between 
each pair of science exposures in order to allow for dispersion correction
and flatfielding.  Sky subtraction was achieved using 20--25 dedicated sky fibres
per setup,
except in the case of M68, where an offset sky exposure was taken.

\subsection{Data Reduction and Analysis}

\begin{figure}
\includegraphics[width=84mm]{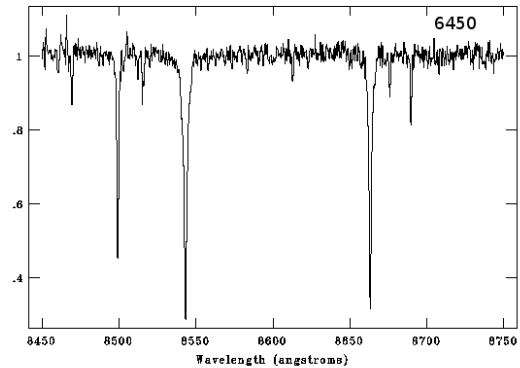}
\caption{Typical spectrum of IC~4499 member RGB star  showing the Ca~II triplet
and many weaker metal lines. Star I.D. 6450 in Table~\ref{tab-stars}} 
\label{fig-spec}
\end{figure}

Data reduction was accomplished using the standard AAOmega reduction software 
\textit{2dfdr drcontrol}.  The
reduction software automatically corrects for CCD bias with blank frames and an overscan bias region. Individual fibre images were traced on the CCD and then dispersion corrected,
wavelength-calibrated spectra were extracted using standard procedures from arc lamp exposures and flat fields.
The M68 sky subtraction was achieved by stacking and averaging the offset sky spectra.
The spectra from separate exposures of the same target were combined using the
IRAF \textit{imcombine} tool. The spectra were normalised by fitting a fifth order
polynomial model to the continuum with the IRAF \textit{onedspec.continuum} task.
Residual cosmic rays in the combined exposures were removed by applying simple clipping.
Given the large sample size, visual inspection of each spectrum was impractical, so
dead fibres, non-stellar objects, and targets with poor signal-to-noise due to 
fibre-centring errors were rejected automatically.
Figure~\ref{fig-spec} shows a typical normalised spectrum.
The spectral resolution achieved was 0.9 \AA\ , with a
typical signal-to-noise ratio in the continuum of 50:1  per pixel.

\subsection{Radial Velocities}
After the data reduction process a total of 36 stars from the calibration 
clusters with velocities from the literature were chosen to be used as radial 
velocity templates. Published references provided online electronic data for
M22 and M4 \citep{peterm22} \citep{peterm4}, which were matched with our 
observations using the ESO SKYCAT software tool. In the case of M68 we used
the published finding charts from \citet{harrism68} to identify the reference
stars.   The chart positions were visually compared with maps made from 
the 2MASS catalogue to match the velocities quoted by Harris to our targets.

Of the 36 available reference stars, 19 had excellent signal to noise ratio, no cosmic rays and
 good sky subtraction residuals. Originally only these 19 reference spectra were employed, but it subsequently proved statistically advantageous to use all available reference spectra to reduce the standard error in the mean of the 36 cross correlations.
Fifteen stars from M4, twelve from M22 and nine from M68 provided a representative sample of the reference clusters. The IRAF task \textit{fxcor} was used to calculate the velocities of the IC~4499 stars by cross correlation with the set of reference spectra \citep{ton79}. The normalised continuum level was subtracted and a Gaussian fitted to the cross correlation to establish the velocity.

The velocity of our target stars was derived from a weighted average of cross-correlation velocities from the
individual template stars.  The average was constructed after automatic rejection of templates that gave large
velocity errors, using a Grubb test.  The velocities based on each remaining template were then averaged,
with weighting based on the cross-correlation errors.

\begin{figure}
\includegraphics[width=84mm]{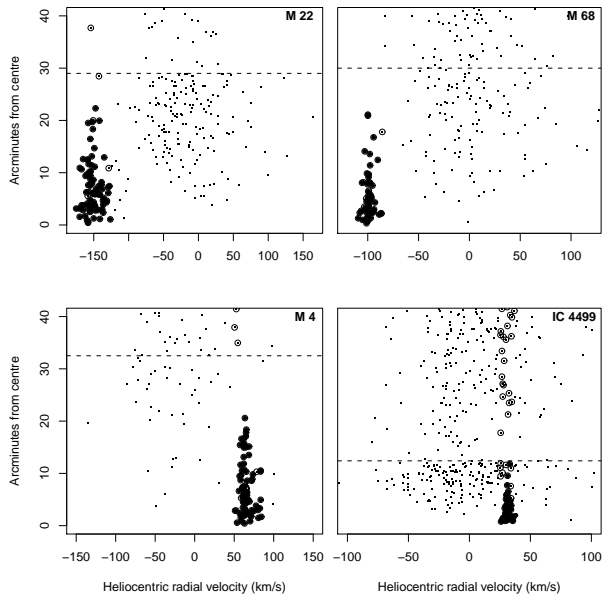}
\caption{Velocities and distances from the cluster centres.  The tidal radii are shown by dashed lines.
Stars within the tidal radius that survived a radial velocity and metallicity clieaning are shown as solid circles . Open circles mark stars with a radial velocity that matches the 
cluster but fall outside the tidal radius, or have Ca~II equivalent widths much different from the cluster members.}
\label{fig-vd}
\end{figure}

Stars were defined as members of the cluster using three parameters. Firstly by distance from
the cluster centre, stars within the tidal radius \citep[catalogued by][]{harris96} were selected. 
Secondly, stars were selected around velocity overdensities. In the case of the calibration clusters
these velocity distributions were located as expected according to previous studies. Stars that
appeared to be normally distributed about these mean values were selected as probable members.  
Figure~\ref{fig-vd} shows the low velocity-dispersion distributions between the
cluster centre and the tidal radius from which stars were selected. Some contamination of the sample from field stars with similar velocities is expected, although this is small for the calibration clusters. 
Stars
were finally selected based on the measured equivalent widths of the three CaII triplet lines
as described in the next section. Apertures were rejected in cases of low S/N, contamination by cosmic rays or other artifacts. These features resulted in odd equivalent width measurements.

\begin{figure}
\includegraphics[width=84mm]{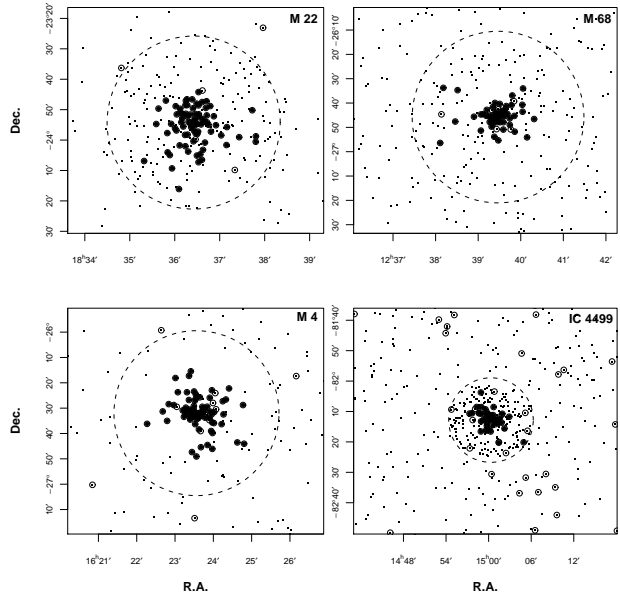}
\caption{Cluster member map. Solid circles are cluster members. Open circles with similar velocities were rejected for lying outside the tidal radius, (dashed line), as metallicity outliers, or for contaminated spectra. }
\label{fig-sel}
\end{figure}

The range of velocities in the disk, outer bulge, and halo towards IC~4499 is quite
large, and there are several field stars projected within the tidal radius.
All velocities have been translated to the heliocentric reference frame within \textit{fxcor}, based on
 the time and date of the observations. The mean cluster velocities
 are given in Table~\ref{tab-res}.  The velocities are in good agreement with literature references
for M22 and M68, but our mean is 5.2 km/sec (4.7$\sigma$) away for M4; the origin of this difference
is unknown.  The mean heliocentric radial velocity of IC~4499 is 31.5 $\pm$0.4 km/sec.  As seen in 
Figure~\ref{fig-sel}, this is sufficiently different from the bulk of the field star velocities
to allow the cluster to be defined, but there is some overlap.

\subsection{Cluster Rotation}
\begin{figure}
\includegraphics[width=84mm]{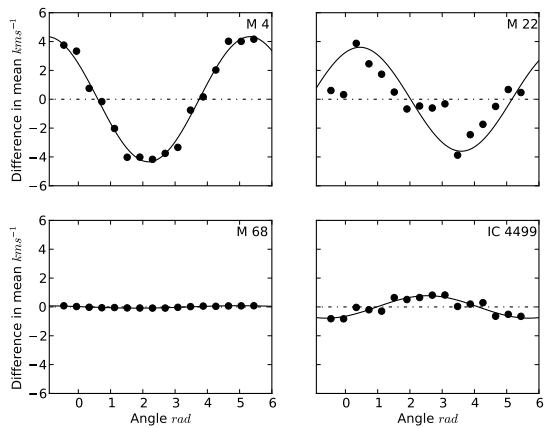}
\caption{Results of cluster rotation searches shows the difference in mean radial velocity
between two halves of the cluster divided by a line orthogonal with the listed position angle.  Position angle is defined 
anticlockwise from East (PA$=$0) through North (PA$=\frac{\pi}{2}$) around the centre of light of the cluster.  The
best-fitting sine curve is shown.}
\label{fig-rot}
\end{figure}

\citet{lane09} find a rotational signature in M22 and a suggestion of one in M68. They also find rotation in M4 \citep{lane10}. While we have a much smaller sample of stars, we can also look for such a signature.
 We employ \citet{lane09}'s method to look for signatures of rotation in M4, M22 and M68 as a check, 
and then in IC~4499.  The position angle of the cluster rotation axis
is not known a priori, so a search of parameter space is made to see if the cluster
radial velocities are consistent with rotation around an arbitrary axis.
For a given trial position angle we divide the cluster in half along a line $90^\circ$ away and
 compare the mean velocity in each
 half of the cluster. We step around the cluster in postion angle increments of $22.5^\circ$.  The asymmetric sky distributions of the samples,(see Figure~\ref{fig-sel}), alias with bin sizes and angular location adding to uncertainty. The differences in
mean radial velocity
 between the cluster halves at each position angle are plotted in Figure~\ref{fig-rot}. 

 We agree with \citet{lane09} on
 the rotation amplitude in M22: we find a line of sight rotational value of 1.8 $\pm0.7$ km/s and the axis of rotation approximately
 North-South, at $114^{\circ}\pm18^{\circ}$, where they found $1.5 \pm0.4$ km/s, approximately North-South. M4 shows
 amplitude $2.1 \pm0.4$ km/s and axis roughly North East-South West at $30^{\circ}\pm12^{\circ}$, about double the result of \citet{lane10} who obtain 0.9 $\pm0.1$ km/s at an angle  of $70^{\circ}$ as do \citet{peterm4}.
 There is no evidence for
 cluster rotation in the data for M68 and IC~449 above the error level of 0.4 km/s.   Our 
M22 and M4 data show that we are sensitive to rotation velocities down to at least $\approx$1 km/s, and this must therefore
be a strict upper limit to the rotation of IC~4499.  No correction for rotational velocity in IC~4499 is necessary when calculating velocity dispersion in the following section.

\subsection{Virial Mass and Mass to Light Ratio}

In order to estimate the cluster mass we need to assume a model for
the cluster gravitational potential and use the central velocity
dispersion, $\sigma_{0}$, and the virial theorem.  Following
\citet{lane10}, a Plummer-type spherical model for the cluster mass
distribution is used, \citep{plu11}. Assuming isotropic velocities one
can calculate a mass using the central velocity dispersion
$\sigma_{0}$, where,
\begin{displaymath}
M = \frac{64\sigma_{0}^2{{R}}}{3\pi\mathcal{G}}
\end{displaymath}
where $R$ is the half-light, or scale, radius and $\mathcal{G}$ is the
gravitational constant.  

To estimate $\sigma_{0}$, \citet{lane10} first bin the velocities by
radius, then use a Markov Chain Monte Carlo (MCMC) technique to estimate dispersions within the bins and subsequently
fit a Plummer model. We have taken a different course, preferring not
to bin the velocities, but choosing instead to fit the Plummer model
to the individual data points.  We assume that the individual
observations are Normally distributed 
\begin{displaymath}
  v_{i} \sim \operatorname{N}\left ( \mu,\sigma^{2}(r) \right )
\end{displaymath}
where the line of sight velocity dispersion $\sigma(r)$ is determined
by a Plummer model 
\begin{displaymath}
\sigma^{2}(r) = \frac{\sigma_{0}^{2}}{\sqrt{1 + (r/R)^{2}}}.
\end{displaymath}
Here $\sigma_{0}$, the central velocity dispersion, is the main
parameter of interest, $R$ the half light radius and $\mu$ the
systemic mean cluster velocity.  Assumptions are made about the initial distributions of parameters, an improper uniform prior for
$\mu$, and weakly informative Normal priors for $R$ and $\sigma_{0}$. We then fit the model 
by MCMC using a Metropolis within Gibbs algorithm \citep{gil98}.
There were $26\times10^{3}$ samples drawn, with the first $6\times10^{3}$ discarded as `burn-in', an initial period where the Markov Chain explores parameter space.

The median value of  the distribution  of $\sigma_{0}$ samples is 2.5 $\pm$0.5 km/s. The median is used as an estimator as the distribution is skewed toward higher values, because the model has a lower bound for central velocity dispersion at zero, but no upper bound. Velocity dispersion has not been constrained to zero at the tidal radius as in a King model \citep{king66}. The mean value is thus slightly higher at 2.6 km/s.
The value of $\mu$, the cluster mean systemic velocity  from MCMC simulation is 31.5 km/s and agrees with the classical sample mean estimator, the sum of velocities divided by the number of samples.
The value of the half light radius from simulation is  $102 \pm18^{\prime\prime}$ and agrees within error with the starting reference value of $107 \pm19^{\prime\prime}$ \citep{trag93}.

\begin{figure}
\includegraphics[width=84mm]{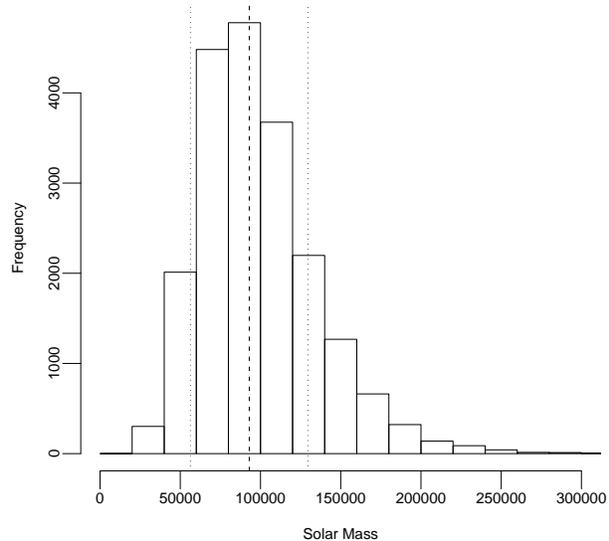}
\caption{Distribution of Markov Chain Monte Carlo simulations of mass, based on a central velocity dispersion model. Standard deviation $37\times10^{3} M_{\odot}$  is shown around the median value, $93\times10^{3} M_{\odot}$. }
\label{fig-cvd}
\end{figure}

The  distribution of cluster mass, a function of the velocity dispersion samples from MCMC simulation, is shown in Figure~\ref{fig-cvd}.  
The median mass is  $93 \pm37\times10^{3} M_{\odot}$ where the error is 1$\sigma$. \citet{mcl05} also estimate a mass for IC~4499 using a power law model, as well as King and Wilson models, fitted to the light distribution of the cluster.
They obtain mass estimates of $125-138\times10^{3} M_{\odot}$ for IC~4499 and central velocity dispersions of $2.88-2.96$ km/s. This spectroscopic  study finds a lower value but  agrees with the photometry-based results within errors.
For an absolute magnitude $M_V = -7.33$, we estimate a mass to light ratio of 1.3 $\pm$0.5 $M/L_{V}$ in solar units. Our lower mass gives a smaller value than \citet{mcl05} who estimate 1.874. This $M/L_{V}$ ratio is similar to other
globulars and indicates that there is not a significant dark matter component to the cluster.

\subsection{Equivalent Widths and Metallicities}

The cluster samples are each assumed to represent a single stellar population. The sample does not
include any stars above the RGB tip, where lower surface gravity results in the line width being more sensitive
to metallicity \citep{caIIsyn}, or low temperature M stars where line width responds more to effective
temperature resulting in lower values, and TiO bands confuse the interpretation.

\begin{figure}
\includegraphics[width=84mm]{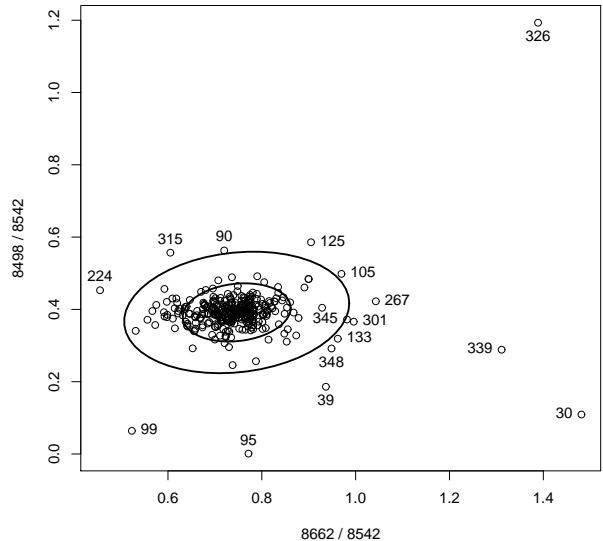}
\caption{Plot of line width ratios of apertures in an IC~4499 field, with 50\% and 95\% confidence contour. Outlying points  were considered statistically unlikely, and rejected aperture numbers are shown . }
\label{fig-ewerr}
\end{figure}

Low signal-to-noise spectra in which one or more of the Ca~II triplet lines were badly distorted
were rejected.  As a diagnostic we compared the ratios of each line with respect to the others.
A line with too large or small a value with respect to the others indicates a problem with the data
or the line fitting results. Spectra with odd line ratios were rejected from further analysis. In
Figure~\ref{fig-ewerr} the ratios are plotted for an IC~4499 cluster sample to identify outliers.

A wavelength range is chosen in the spectrum that encompasses the line feature and a 
representative portion of the normalised continuum. The line and continuum
 bandpasses are taken as defined in \citet{arm91}.
The sum of a Gaussian and a Lorentzian, a Penny function, is fitted to the line profiles using the same FORTRAN code as \citet{cole2004}, which is a modified version of the code of \citet{arm1986}, to give an equivalent width for each line. The Penny function has been shown to be a better approximation for high metallicity, high resolution spectra \citep{war09}. Model atmospheres of late-type giants indicate the wings are more sensitive to  metallicity than other parameters such as  surface gravity and effective temperature \citep{drake90}.

The equivalent widths of the three triplet lines are summed to give the CaII index. Some authors sum the two strongest lines for low signal to noise data or low resolution spectra \citep{rutl97b}. Here, having a high signal to noise ratio, the sum of three lines is taken to give the full equivalent width $\Sigma$W. Results for $\Sigma$W in the sample  stars in IC~4499 are shown in Table~\ref{tab-stars}.

Next a reduced equivalent width $W^{\prime}$ is derived in which the linear dependence on the magnitude height on the RGB is taken into account. This \emph{magnitude} parameter represents the effects of the effective temperature and stellar surface gravity on the line strengths \citep{arm91}. Because red giants
lie along a narrow sequence in the luminosity (surface gravity) vs. temperature plane,
T$_{eff}$ and log$g$ are correlated with each other and their influence on $\Sigma$W
can be calibrated out using a single observable. Colour and absolute magnitude have
both been used in the past to create the index W$^{\prime}$, but the most robust method
in the presence of distance and reddening uncertainties is to use an expression relating
the magnitude of the target star to the mean magnitude of the horizontal branch (or 
red clump) of its parent population \citep{rutl97b}.

Owing to the availability and homogeneity of 2MASS near-infrared magnitudes, we adopt the procedure of
\citet{war09} and use the K-band magnitude, K$-$K$_{HB}$ to derive $W^{\prime}$.  \citet{war09}
defined W$^{\prime}$ $=$ $\Sigma$W $+$0.45(K$-$K$_{HB}$); because it is uncommon to use the K band
in this procedure, the slope is not as well-determined as that in V or I, so we rederive the
relationship as a consistency check.

The mean value of the RR~Lyrae variable magnitudes is used to define the horizontal branch K-magnitude, 12.21 for M22, 11.13 for M4 and 15.97 for IC~4499.
K-magnitudes for RR~Lyrae variables in IC4499 are found in \citet{storm04} and define the
magnitude of the horizontal branch.
There are K-magnitudes for a few RR~Lyrae stars in M22 \citep{kaluzny}, and several for M4 \citep{liu}.
Many more M4 variables are listed in \citet{long}. RR~Lyrae variables in M22 and M4 were identified from those catalogued in \citet{clement} and these were astrometrically correlated with 2MASS objects to obtain K magnitudes.
M68 horizontal branch K-magnitude of 14.4 is referenced from \citet{ferraro00} and \citet{dallora}.
For each cluster K$_{HB}$ is taken to be constant and each star in the 2MASS PSC has a unique K$-$K$_{HB}$.

\begin{figure}
\includegraphics[width=84mm]{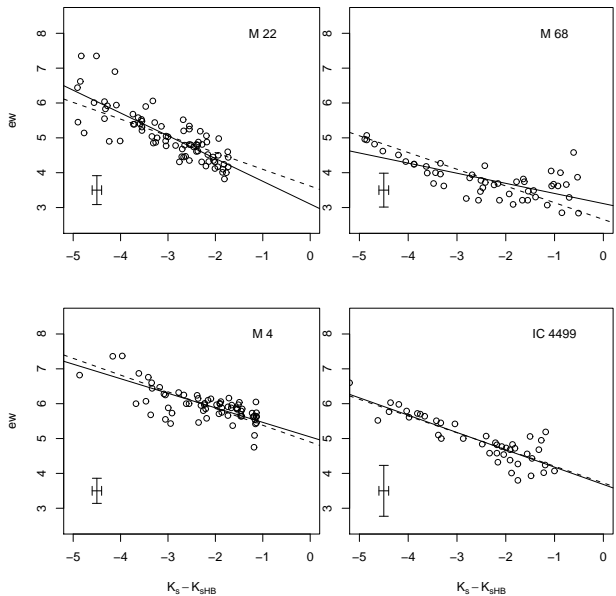}
\caption{K$-$K$_{HB}$ vs.\ $\Sigma$W for our clusters.  The  dashed line shows the average slope from \citet{war09},
and the  solid lines give the best-fit slope for each individual cluster based on our data. Typical errorbars 
are shown in the lower left of each panel.  The intercept of the relation defines the \emph{reduced} equivalent
width W$^{\prime}$ for each cluster.}
\label{fig-ew}
\end{figure}

We plot the relative magnitude $K-K_{HB}$ against the equivalent width $\Sigma$W in
Figure~\ref{fig-ew}. The slopes of the
lines $\beta_{K}$ range from 0.29 $\leq$ $\beta_K$ $\leq$ 0.65, with a mean value of 0.47 $\pm$0.08 \AA/mag.
This agrees well with the value of $\beta_{K} = 0.48 \pm0.01$ found by \citet{war09}, who have 22 clusters,
open and globular, in their sample. There is no strong reason to suspect variation in $\beta$
for a globular cluster-only sample \citep{rutl97b}, so we adopt the better-determined value
$\beta_K$ = 0.48 $\pm$0.01 from \citet{war09}.  The
\emph{reduced} equivalent width $W^{\prime}$, is the intercept of this linear model.  W$^{\prime}$ should be
a constant for each cluster that only depends on metallicity.  Fits to our four clusters and the literature slope are
shown in Figure~\ref{fig-ew} and the values of W$^{\prime}$ listed in Table~\ref{tab-res}.

$W^{\prime}$ is related linearly to [Fe/H] on the \citet{cg97} scale; we follow \citet{war09} who arrived
at the following relation for transforming to metallicity:
\begin{displaymath}
 [Fe/H]=(-2.738 \pm 0.063) + (0.330 \pm 0.009)W^{\prime}
\end{displaymath}

The values of [Fe/H] derived from this relationship are given in Table~\ref{tab-res}.  They agree with
the literature values to better than 1$\sigma$, as expected.  As emphasised by \citet{cole2004,war09},
these values are specific to the \citet{cg97} metallicity scale.  Recalibration to ZW84, the scale
of \citet{kra03}, or any other metallicity scale of choice may be achieved using the W$^{\prime}$
values, which give the correct relative metallicity ranking of the clusters no matter the specific
W$^{\prime}$-[Fe/H] conversion adopted.  

\section{The Velocity and Metallicity of IC~4499}

Identifications, positions, velocities and equivalent widths for individual IC~4499 stars are given in Table~\ref{tab-stars}.
The metallicity of IC~4499 is very similar to the mean metallicity of M22, [Fe/H] = $-$1.52 $\pm$0.12, and 
the radial velocity is v$_r$ = 31.5 $\pm$0.4 km/s.  This is the first published spectroscopic
metallicity measurement for the cluster based on more than just a few stars.  Previous
estimates for the radial velocity varied widely and are difficult to properly assess.

\citet{perk} derived a spectroscopic metallicity of [Fe/H] = $-$1.33 $\pm0.3$ from three RR~Lyrae variables 
in IC~4499. ZW84 revise this figure by recalibrating to the scale of \citet{frogel} and quote 
[Fe/H] = $-$1.5 $\pm0.3$.   \citet{ferraro95} found this value to be too high, and adopted $-$1.75
based on the CMD morphology, primarily HB type and RGB colour.  Later studies of the RR~Lyrae population
\citep[e.g.,][]{walker96} found no inconsistencies with this value, and cite an unpublished study by
R. Cannon finding [Fe/H] = $-$1.65 in support.  Our value of [Fe/H] = $-$1.52 $\pm$0.12 on the \citet{cg97}
scale translates to [Fe/H] = $-$1.74 $\pm$0.10 on the ZW scale.  M22, a cluster with very similar 
W$^{\prime}$, has [Fe/H] = $-$1.75 according to \citet{cg97}, and [Fe/H] = $-$1.9 on the scale
of \citet{kra03}, according to \citet{dac09}.  The latter paper also finds strong evidence for
an internal spread of metallicities in M22 of up to 0.5~dex, so more detailed comparison to M22
may only serve to confuse the picture.
However, we can confirm that the RR~Lyrae-based 
result from \citet{perk} for IC~4499 is too metal-rich, and the CMD results are robust.

The radial velocity measurement is relatively unexceptional, as a wide range of velocities are expected
towards the 4th quadrant of the Galaxy.  The Besan\c{c}on model Galaxy \citep{rob03} shows that radial
velocities toward IC~4499 have a broad maximum around $-$15 km/s, with FWHM $\approx$60 km/s. If only
the stars with [Fe/H] $\leq$ $-$1 are considered, the mean radial velocity is expected to be $\approx$25
km/s, with a very broad distribution: the FWHM of metal-weak stars in this direction is $\approx$110 km/s,
and with tails reaching to $-$150 $\lesssim$ v$_r$ $\lesssim$ $+$350 km/s.  IC~4499 thus lies near the
peak of the expected radial velocity distribution of halo stars in this direction.
We are in disagreement with the average velocity of 3  RR~Lyrae stars,
$-$50 km/s, reported by \citet{perk}.  Other velocity measurements are scarce; \citet{mon05} place
IC~4499 in their Fig.~11 with v$_r$ = 0, without attribution.  Similarly, \citet{fus95} give the
cluster a  galactocentric radial velocity v$_{r,GC}$ $\approx$ $-$130~km/s, also without citing a source.  
The \citet{fus95} value is not far from the measured value of v$_r$ if we account for the solar motion;
we find v$_{r,GC}$ = $-$140~km/s.  

%

\subsection{Is IC~4499 Unusual?}

IC~4499 has an exceptionally high frequency of RR~Lyrae stars \citep[e.g.,][and references therein]{walker96}, 
and has been proposed to be 2--4~Gyr younger than the average of metal-poor clusters \citep{ferraro95}.  
It is also a candidate to belong to halo substructures \citep{fus95}, including the possibility of 
membership in the Monoceros tidal stream if that structure is due to the dissolution of a dwarf galaxy
in the tidal field of the Milky Way \citep{mon05}.  These suggestions hint towards the idea that the
HB morphology can be connected to some combination of age and/or detailed elemental abundance ratios
(e.g., differences in [$\alpha$/Fe]).   A further clue may be in the fact that the cluster is of
Oosterhoff type I, that is, the  RR~Lyrae stars have periods $\lesssim$0.6 d.  

It is well-known that the Oosterhoff type of a cluster is related to its metallicity 
\citep[e.g.,][]{walker96}, but the relation is not a straightforward one.
In general, the period of the variation increases with decreasing metallicity,
but several
clusters have been found that appear to break the rules. NGC 6388 and NGC 6441
\citep{pritzl00}, are metal-rich clusters displaying properties of both Oosterhoff
types. 

\citet{sand93} noted that there are very few variables in clusters with
 $-$1.7 $\leq$ [Fe/H] $\leq$ $-$1.9 on the ZW84 scale, and that the few known
 RR~Lyrae stars present have longer than expected periods.   However, he assumed
[Fe/H] = $-$1.5 for IC~4499, where it should have a [Fe/H] = $-$1.75 on the
ZW84 scale.  Clusters of similar metallicity indeed tend to have much smaller
specific frequencies of  RR~Lyrae stars \citep[e.g., M22 has S$_{RR}$ = 7.2,][]{harris96}.
This is likely because most of the HB stars at this metallicity begin their
lifetimes well to the blue of the RR~Lyrae instability strip \citep{lee90}.
The extremely high specific frequency of  RR~Lyrae stars at the metallicity of 
IC~4499 suggests a larger than average stellar mass at the zero-age horizontal branch
(ZAHB).  As noted by \citet{walker96}, this could be indicative of a younger than
average age for the cluster, or it could suggest a smaller-than-average amount
of mass-loss along the cluster RGB; \citet{sand93} already suggested that a
smaller-than-average {\it dispersion} in mass-loss was necessary to reproduce
the colour extent of IC~4499's HB.  IC~4499's lower-than-average central density
compared to clusters like M3 may be related to its RGB mass-loss behaviour.
It is also possible that variations in the
detailed elemental abundances, such as [O/Fe], play a role in determining the HB 
morphology. 
 
We find M22 to have similar average metallicity to IC~4499; it has few  RR~Lyrae stars,
a blue HB, and a higher central density, and is an OoII type cluster.  This
makes the 2 clusters something like a classical ``second parameter'' pair like
M3 and M2, both with [Fe/H] $\approx$ $-$1.6 on the CG97 scale.  \citet{lee99}
have proposed that there is an age difference between the two clusters of 
$\approx$2~Gyr, in accordance with the arguments in, e.g., \citet{lee92}.
This is similar to the argument in \citet{ferraro95} that IC~4499 is
$\approx$2--4~Gyr younger than similar-metallicity halo clusters.
However the picture is complicated here because of the existence of a
significant range of abundances in M22 \citep{dac09}.
IC~4499 appears to have a slightly unusual \citet{lee89} HB type
for its metallicity, but a younger than average age cannot definitely be
stated to be the cause.  According to the models in \citet{lee92}, an 
age difference of $\lesssim$1~Gyr compared to M3 would be required
to account for the relatively red HB morphology; the difference would be smaller
if smaller-than-average mass-loss is adopted.  Comparing to the HB types
of other candidate ``young'' globulars, IC~4499 is likely to be significantly
older than Rup~106, and some of the outer halo clusters like Pal~4 and Eridanus.
This complicates the suggestion in \citet{fus95} for a common
origin shared between Rup~106 and IC~4499.  

\citet{sal02} found that
IC~4499 was nearly coeval with other intermediate-metallicity clusters
However, they assumed
an incorrect metallicity, [Fe/H] = $-$1.26 (CG97), and the comparison
should be redone using the more accurate value [Fe/H] = $-$1.52 $\pm$0.12.
\citet{sal02} conclude that all clusters with [Fe/H] $\leq$ $-$1.2
appear to be coeval within errors, at an age of $\approx$12~Gyr.
\citet{forbes10} argue that there is a break in the age-metallicity
relation for galactic globular clusters at [Fe/H] $\approx$ $-$1.5.
 While there exists a class of old clusters at higher metallicity, there
 appears to be a group of young clusters with metallicities above the
break point which they identify as accreted ``young halo'' clusters. 
 Like \citet{sal02}, \citet{forbes10} find those 
classified as ``old halo'' are roughly coeval at $\approx$12.8~Gyr.
If a reanalysis of the cluster CMD is made, using the new spectroscopic
metallicity, that still suggests an age difference relative to the bulk
of halo globulars, then the conclusion of \citet{sal02} would be challenged
and it would suggest that IC~4499 belongs to the ``young'' group of \citet{forbes10}
clusters.

\citet{cor07} hypothesise that halo objects are divided into two main classes,
 with the outer halo having lower metallicity and odd orbits suggesting accretion
 from low-mass dwarf galaxies, while the inner halo is higher metallicity
 and galactic in origin.  IC~4499 has a smaller galactocentric distance than
typical outer halo clusters, but its location where models predict an extension 
of the Monoceros tidal stream \citep{mon05} may strengthen the idea that it has
an extragalactic origin.  The evidence for membership in the Monoceros stream
to date has been based solely on its position within a modeled extension of the 
stream.  At the location of IC~4499, these models predict a radial velocity
that ranges between $-$60 km/s $\leq$ v$_r$ $\lesssim$ 100 km/s, which has nearly
complete overlap with the standard Galactic halo model for this sightline
\citep{rob03}.  An interesting feature of the \citet{mon05} model is that
as in \citet{fus95}, Rup~106 and IC~4499 are suggested to be members of a single
dynamically-related feature, but in the Monoceros stream model Rup~106 belongs
to the trailing side of the tidal stream, while IC~4499 is a member of 
the leading stream.  Both clusters have drifted a large distance from 
their progenitor: nearly 360$^{\circ}$ in the case of IC~4499, more
than a complete wrap for Rup~106, and their apparent positioning as
neighbours along a single great circle is coincidental.

Using the methodology of \citet{van93}
to classify the orbital parameters of IC~4499, the cluster is likely to
be in a prograde orbit that is of a ``plunging'' type.  However, the
cluster lies near the limit for circular orbits, suggesting that there is
a relatively high likelihood that it is on a mildly eccentric orbit.  Placing 
the cluster in context, it appears quite normal for its galactocentric distance
and metallicity, and membership in a tidal stream is not needed to explain
its radial velocity.  Because the predicted radial velocity of the Monoceros tidal stream is consistent with
the expectations for the general field, further information is necessary before
IC~4499 can definitely be assigned membership in a stream, or be inferred to have been
accreted into the halo from a dwarf galaxy.   Two observables that could help 
discriminate between models are the cluster proper motion and the detailed abundance 
ratios of the member stars. 

The Monoceros stream model that matches the position and
distance of IC~4499 predicts a range of proper motions in galactic coordinates,
($-$4, $-$1) $\lesssim$ ($\mu_l$, $\mu_b$) $\lesssim$ ($-$2, $+$2) mas/yr.  On the
other hand, the Besan\c{c}on model suggests that most late-type halo stars at
IC~4499's location will have proper motions of ($\mu_l$, $\mu_b$) $\approx$
($-$5$\pm$5, 0$\pm$5) mas/yr.  From this it can be seen that a proper motion
in the appropriate range for stream membership is not sufficient to ensure 
membership, since halo stars overlap in both components (although less so in
$\mu_l$).  Proper motions could make a strong negative test in that the cluster
could be excluded from stream membership via this measurement.

Detailed abundance ratios are a stronger test, because of the consistency
of elemental abundance ratios among field and cluster halo stars \citep[e.g.,][]{ful00,ful02} and
the strong anomalies seen in dwarf spheroidal galaxy field stars \citep[e.g.,][]{shet03,mcw05,chou10} and
clusters \citep[e.g.,][and references therein]{bell08,car10}.
 In particular the [$\alpha$/Fe] vs.\ [Fe/H] trend and ratios
of $s$- and $r$-process elements can give strong clues to the past star-formation
history, initial mass function sampling, and loss of metals from a stellar system
\citep{tol03,venn04}. Because IC~4499 is not a very massive cluster, it is expected to be
chemically homogeneous, and high-resolution spectra of just a few stars should
suffice to begin characterisation of its nucleosynthetic history.

\section{Conclusions}

We obtained near-infrared spectra of 636 red giants in and around the
RR~Lyrae-rich,
extreme-southern globular cluster IC~4499.  From spectra including the
calcium triplet, we measured 
radial velocities by cross-correlation with template stars in
well-studied globular clusters M68, M22, and M4.
By combining the CaT equivalent widths with 2MASS K$_s$ magnitudes
we derived metallicities on the \citet{cg97} scale following the
methodology of \citet{war09}.   The relationship in our data between
CaT equivalent widths, K$_s$, and [Fe/H] is in good agreement with
the work of \citet{war09}.
Our velocity and metallicity results
for the comparison clusters agree well with literature values.
43 stars were found to be probable cluster members based on radial 
velocity association, culled by metallicity to alleviate the strong
foreground contamination.

The heliocentric radial velocity of IC~4499 is v$_r$ = 31.5 $\pm$0.4
km/s.  The velocity is typical of halo objects along this sightline,
and also does not rule out membership in a tidal stream as proposed
by \citet{mon05}.  The most powerful tests of stream membership,
proper motion and detailed elemental abundance ratios, are not
yet available for this cluster.  Like many proposed associations
\citep[e.g.,][]{pia08} the status of IC~4499 is undecided.

The metallicity of IC~4499 is [Fe/H] = $-$1.52 $\pm$0.12 on the
scale of \citet{cg97}, which translates to $-$1.74 $\pm$0.10 on
the \citet{zinn} scale.  This agrees with photometric estimates
from the cluster CMD and unpublished work by R. Cannon (1992),
but disagrees with the earlier studies of RR~Lyrae stars \citep{perk}.
To the extent that studies of the relative ages of globular clusters
\citep[e.g.,][]{sal02} and of the Oosterhoof RR~Lyrae period-metallicity
relation \citep[e.g.,][]{sand93} incorrectly relied on overestimates
of the cluster metallicity, the role of IC~4499 in these studies should
be reassessed.  
If age is the dominant contributor to the second-parameter effect
\citep{lee92}, then the evidence for a young age \citep{ferraro95}
for IC~4499 is weak, based on its intermediate HB type.  The cluster
is slightly metal-poor compared to most OoI clusters.

We follow the approach of \citet{lane09,lane10} to search for evidence
of rotation in IC~4499.  We confirm their results in M22 and M4, although
the signal is noisy because we have measured less than half the number of
stars.  We are unable to detect evidence for rotation in IC~4499, which
puts an upper limit of $\approx$1~km/s on the net cluster rotation.

We model the velocity dispersion of the cluster using a Plummer potential,
finding the best-fit cluster parameters using a Markov Chain Monte Carlo
simulation.  The most likely central velocity dispersion is
$\sigma_0$ = 2.5 $\pm$0.5~km/s.  Using the Plummer model this translates
to a cluster dynamical mass of 93 $\pm$37$\times$10$^{3}$ M$_{\odot}$.
This is in agreement with fits to the light profile by \citet{mcl05},
and using their photometry implies a mass-to-light ratio M/L$_V$ = 1.3
in solar units; this result is quite normal for a globular cluster
\citep[e.g.,][]{trag93,lane10} and says that no dark matter component
is needed to explain the cluster dynamics.

\section*{Acknowledgments}Travel support for AAC was provided by the Anglo-Australian
Observatory (AAO). The AAO is funded by the British and Australian governments.
WJH acknowledges the support of the Grote Reber Foundation.
AAC would like to thank AAT support astronomer Paul Dobbie and
night assistant Winston Campbell for their assistance during the observing run. 
This publication makes use of data products from the Two Micron All Sky Survey,
which is a joint project of the University of Massachusetts and IPAC/Caltech,
funded by NASA and the NSF. This research has made use of the WEBDA database,
operated at the Institute for Astronomy of the University of Vienna.
IRAF is distributed by the National Optical Astronomy Observatories.
The European Organisation for Astronomical Research in the Southern Hemisphere (ESO) maintain and distribute SKYCAT.
Thanks to Dr. Simon Wotherspoon of the University of Tasmania for scripting the MCMC algorithm.

\bibliographystyle{mn2e}
\bibliography{ic4499}

\begin{thebibliography}{}

\bibitem[\protect\citeauthoryear{{Armandroff} \& {Da Costa}}{{Armandroff} \&
  {Da Costa}}{1986}]{arm1986}
{Armandroff} T.~E.,  {Da Costa} G.~S.,  1986, \aj, 92, 777

\bibitem[\protect\citeauthoryear{{Armandroff} \& {Da Costa}}{{Armandroff} \&
  {Da Costa}}{1991}]{arm91}
{Armandroff} T.~E.,  {Da Costa} G.~S.,  1991, \aj, 101, 1329

\bibitem[\protect\citeauthoryear{{Armandroff} \& {Zinn}}{{Armandroff} \&
  {Zinn}}{1988}]{arm1988}
{Armandroff} T.~E.,  {Zinn} R.,  1988, \aj, 96, 92

\bibitem[\protect\citeauthoryear{{Battaglia}, {Irwin}, {Tolstoy}, {Hill},
  {Helmi}, {Letarte} \& {Jablonka}}{{Battaglia} et~al.}{2008}]{bat08}
{Battaglia} G.,  {Irwin} M.,  {Tolstoy} E.,  {Hill} V.,  {Helmi} A.,  {Letarte}
  B.,    {Jablonka} P.,  2008, \mnras, 383, 183

\bibitem[\protect\citeauthoryear{{Bellazzini}, {Ibata}, {Chapman}, {Mackey},
  {Monaco}, {Irwin}, {Martin}, {Lewis} \& {Dalessandro}}{{Bellazzini}
  et~al.}{2008}]{bell08}
{Bellazzini} M.,  {Ibata} R.~A.,  {Chapman} S.~C.,  {Mackey} A.~D.,  {Monaco}
  L.,  {Irwin} M.~J.,  {Martin} N.~F.,  {Lewis} G.~F.,    {Dalessandro} E.,
  2008, \aj, 136, 1147

\bibitem[\protect\citeauthoryear{{Carollo}, {Beers}, {Lee}, {Chiba}, {Norris},
  {Wilhelm}, {Sivarani}, {Marsteller}, {Munn}, {Bailer-Jones}, {Fiorentin} \&
  {York}}{{Carollo} et~al.}{2007}]{cor07}
{Carollo} D.,  {Beers} T.~C.,  {Lee} Y.~S.,  {Chiba} M.,  {Norris} J.~E.,
  {Wilhelm} R.,  {Sivarani} T.,  {Marsteller} B.,  {Munn} J.~A.,
  {Bailer-Jones} C.~A.~L.,  {Fiorentin} P.~R.,    {York} D.~G.,  2007, \nat,
  450, 1020

\bibitem[\protect\citeauthoryear{{Carretta}, {Bragaglia}, {Gratton},
  {Lucatello}, {Bellazzini}, {Catanzaro}, {Leone}, {Momany}, {Piotto} \&
  {D'Orazi}}{{Carretta} et~al.}{2010}]{car10}
{Carretta} E.,  {Bragaglia} A.,  {Gratton} R.~G.,  {Lucatello} S.,
  {Bellazzini} M.,  {Catanzaro} G.,  {Leone} F.,  {Momany} Y.,  {Piotto} G.,
  {D'Orazi} V.,  2010, ArXiv e-prints

\bibitem[\protect\citeauthoryear{{Carretta} \& {Gratton}}{{Carretta} \&
  {Gratton}}{1997}]{cg97}
{Carretta} E.,  {Gratton} R.~G.,  1997, \aaps, 121, 95

\bibitem[\protect\citeauthoryear{{Chou}, {Cunha}, {Majewski}, {Smith},
  {Patterson}, {Mart{\'{\i}}nez-Delgado} \& {Geisler}}{{Chou}
  et~al.}{2010}]{chou10}
{Chou} M.,  {Cunha} K.,  {Majewski} S.~R.,  {Smith} V.~V.,  {Patterson} R.~J.,
  {Mart{\'{\i}}nez-Delgado} D.,    {Geisler} D.,  2010, \apj, 708, 1290

\bibitem[\protect\citeauthoryear{{Clement}, {Muzzin}, {Dufton}, {Ponnampalam},
  {Wang}, {Burford}, {Richardson}, {Rosebery}, {Rowe} \& {Hogg}}{{Clement}
  et~al.}{2001}]{clement}
{Clement} C.~M.,  {Muzzin} A.,  {Dufton} Q.,  {Ponnampalam} T.,  {Wang} J.,
  {Burford} J.,  {Richardson} A.,  {Rosebery} T.,  {Rowe} J.,    {Hogg} H.~S.,
  2001, \aj, 122, 2587

\bibitem[\protect\citeauthoryear{{Clement} \& {Rowe}}{{Clement} \&
  {Rowe}}{2000}]{clem00}
{Clement} C.~M.,  {Rowe} J.,  2000, \aj, 120, 2579

\bibitem[\protect\citeauthoryear{{Cole}, {Smecker-Hane}, {Tolstoy}, {Bosler} \&
  {Gallagher}}{{Cole} et~al.}{2004}]{cole2004}
{Cole} A.~A.,  {Smecker-Hane} T.~A.,  {Tolstoy} E.,  {Bosler} T.~L.,
  {Gallagher} J.~S.,  2004, \mnras, 347, 367

\bibitem[\protect\citeauthoryear{{Da Costa}, {Held}, {Saviane} \&
  {Gullieuszik}}{{Da Costa} et~al.}{2009}]{dac09}
{Da Costa} G.~S.,  {Held} E.~V.,  {Saviane} I.,    {Gullieuszik} M.,  2009,
  \apj, 705, 1481

\bibitem[\protect\citeauthoryear{{Dall'Ora}, {Bono}, {Storm}, {Caputo},
  {Andreuzzi}, {Marconi}, {Monelli}, {Ripepi}, {Stetson} \& {Testa}}{{Dall'Ora}
  et~al.}{2006}]{dallora}
{Dall'Ora} M.,  {Bono} G.,  {Storm} J.,  {Caputo} F.,  {Andreuzzi} G.,
  {Marconi} G.,  {Monelli} M.,  {Ripepi} V.,  {Stetson} P.~B.,    {Testa} V.,
  2006, Memorie della Societa Astronomica Italiana, 77, 214

\bibitem[\protect\citeauthoryear{{Ferraro}, {Montegriffo}, {Origlia} \& {Fusi
  Pecci}}{{Ferraro} et~al.}{2000}]{ferraro00}
{Ferraro} F.~R.,  {Montegriffo} P.,  {Origlia} L.,    {Fusi Pecci} F.,  2000,
  \aj, 119, 1282

\bibitem[\protect\citeauthoryear{{Ferraro}, {Ferraro}, {Pecci}, {Corsi} \&
  {Buonanno}}{{Ferraro} et~al.}{1995}]{ferraro95}
{Ferraro} I.,  {Ferraro} F.~R.,  {Pecci} F.~F.,  {Corsi} C.~E.,    {Buonanno}
  R.,  1995, \mnras, 275, 1057

\bibitem[\protect\citeauthoryear{{Forbes} \& {Bridges}}{{Forbes} \&
  {Bridges}}{2010}]{forbes10}
{Forbes} D.~A.,  {Bridges} T.,  2010, \mnras, 404, 1203

\bibitem[\protect\citeauthoryear{{Fourcade}, {Laborde} \& {Arias}}{{Fourcade}
  et~al.}{1974}]{fou74}
{Fourcade} C.~R.,  {Laborde} J.~R.,    {Arias} J.~C.,  1974, \aaps, 18, 3

\bibitem[\protect\citeauthoryear{{Frinchaboy}, {Majewski}, {Crane}, {Reid},
  {Rocha-Pinto}, {Phelps}, {Patterson} \& {Mu{\~n}oz}}{{Frinchaboy}
  et~al.}{2004}]{frinch}
{Frinchaboy} P.~M.,  {Majewski} S.~R.,  {Crane} J.~D.,  {Reid} I.~N.,
  {Rocha-Pinto} H.~J.,  {Phelps} R.~L.,  {Patterson} R.~J.,    {Mu{\~n}oz}
  R.~R.,  2004, \apjl, 602, L21

\bibitem[\protect\citeauthoryear{{Frogel}, {Cohen} \& {Persson}}{{Frogel}
  et~al.}{1983}]{frogel}
{Frogel} J.~A.,  {Cohen} J.~G.,    {Persson} S.~E.,  1983, \apj, 275, 773

\bibitem[\protect\citeauthoryear{{Fulbright}}{{Fulbright}}{2000}]{ful00}
{Fulbright} J.~P.,  2000, \aj, 120, 1841

\bibitem[\protect\citeauthoryear{{Fulbright}}{{Fulbright}}{2002}]{ful02}
{Fulbright} J.~P.,  2002, \aj, 123, 404

\bibitem[\protect\citeauthoryear{{Fusi Pecci}, {Bellazzini}, {Cacciari} \&
  {Ferraro}}{{Fusi Pecci} et~al.}{1995}]{fus95}
{Fusi Pecci} F.,  {Bellazzini} M.,  {Cacciari} C.,    {Ferraro} F.~R.,  1995,
  \aj, 110, 1664

\bibitem[\protect\citeauthoryear{{Garcia-Vargas}, {Molla} \&
  {Bressan}}{{Garcia-Vargas} et~al.}{1998}]{caIIsyn}
{Garcia-Vargas} M.~L.,  {Molla} M.,    {Bressan} A.,  1998, \aaps, 130, 513

\bibitem[\protect\citeauthoryear{{Geisler}, {Piatti}, {Claria} \&
  {Minniti}}{{Geisler} et~al.}{1995}]{geisler}
{Geisler} D.,  {Piatti} A.~E.,  {Claria} J.~J.,    {Minniti} D.,  1995, \aj,
  109, 605

\bibitem[\protect\citeauthoryear{{Gilks}, {Richardson} \&
  {Spiegelhalter}}{{Gilks} et~al.}{1998}]{gil98}
{Gilks} W.~R.,  {Richardson} S.,    {Spiegelhalter} D.~J.,  1998, Markov Chain
  Monte Carlo in Practice.
Chapman and Hall, Boca Raton, Florida

\bibitem[\protect\citeauthoryear{{Grocholski}, {Cole}, {Sarajedini}, {Geisler}
  \& {Smith}}{{Grocholski} et~al.}{2006}]{gro06}
{Grocholski} A.~J.,  {Cole} A.~A.,  {Sarajedini} A.,  {Geisler} D.,    {Smith}
  V.~V.,  2006, \aj, 132, 1630

\bibitem[\protect\citeauthoryear{{Harris}}{{Harris}}{1975}]{harrism68}
{Harris} W.~E.,  1975, \apjs, 29, 397

\bibitem[\protect\citeauthoryear{{Harris}}{{Harris}}{1996}]{harris96}
{Harris} W.~E.,  1996, VizieR Online Data Catalog, 7195, 0

\bibitem[\protect\citeauthoryear{{Helmi}, {Navarro}, {Meza}, {Steinmetz} \&
  {Eke}}{{Helmi} et~al.}{2003}]{hel03}
{Helmi} A.,  {Navarro} J.~F.,  {Meza} A.,  {Steinmetz} M.,    {Eke} V.~R.,
  2003, \apjl, 592, L25

\bibitem[\protect\citeauthoryear{{Ibata}, {Irwin}, {Lewis}, {Ferguson} \&
  {Tanvir}}{{Ibata} et~al.}{2003}]{iba03}
{Ibata} R.~A.,  {Irwin} M.~J.,  {Lewis} G.~F.,  {Ferguson} A.~M.~N.,
  {Tanvir} N.,  2003, \mnras, 340, L21

\bibitem[\protect\citeauthoryear{{Kaluzny} \& {Thompson}}{{Kaluzny} \&
  {Thompson}}{2001}]{kaluzny}
{Kaluzny} J.,  {Thompson} I.~B.,  2001, VizieR Online Data Catalog, 337, 30899

\bibitem[\protect\citeauthoryear{{King}}{{King}}{1966}]{king66}
{King} I.~R.,  1966, \aj, 71, 64

\bibitem[\protect\citeauthoryear{{Kraft} \& {Ivans}}{{Kraft} \&
  {Ivans}}{2003}]{kra03}
{Kraft} R.~P.,  {Ivans} I.~I.,  2003, \pasp, 115, 143

\bibitem[\protect\citeauthoryear{{Lane}, {Kiss}, {Lewis}, {Ibata}, {Siebert},
  {Bedding} \& {Sz{\'e}kely}}{{Lane} et~al.}{2009}]{lane09}
{Lane} R.~R.,  {Kiss} L.~L.,  {Lewis} G.~F.,  {Ibata} R.~A.,  {Siebert} A.,
  {Bedding} T.~R.,    {Sz{\'e}kely} P.,  2009, \mnras, 400, 917

\bibitem[\protect\citeauthoryear{{Lane}, {Kiss}, {Lewis}, {Ibata}, {Siebert},
  {Bedding}, {Sz{\'e}kely}, {Balog} \& {Szab{\'o}}}{{Lane}
  et~al.}{2010}]{lane10}
{Lane} R.~R.,  {Kiss} L.~L.,  {Lewis} G.~F.,  {Ibata} R.~A.,  {Siebert} A.,
  {Bedding} T.~R.,  {Sz{\'e}kely} P.,  {Balog} Z.,    {Szab{\'o}} G.~M.,  2010,
  \mnras, pp 799--+

\bibitem[\protect\citeauthoryear{{Lee} \& {Carney}}{{Lee} \&
  {Carney}}{1999}]{lee99}
{Lee} J.,  {Carney} B.~W.,  1999, \aj, 118, 1373

\bibitem[\protect\citeauthoryear{{Lee}}{{Lee}}{1989}]{lee89}
{Lee} Y.,  1989, PhD thesis, Yale Univ., New Haven, CT.

\bibitem[\protect\citeauthoryear{{Lee}}{{Lee}}{1992}]{lee92}
{Lee} Y.,  1992, \pasp, 104, 798

\bibitem[\protect\citeauthoryear{{Lee}, {Demarque} \& {Zinn}}{{Lee}
  et~al.}{1990}]{lee90}
{Lee} Y.,  {Demarque} P.,    {Zinn} R.,  1990, \apj, 350, 155

\bibitem[\protect\citeauthoryear{{Liu} \& {Janes}}{{Liu} \&
  {Janes}}{1990}]{liu}
{Liu} T.,  {Janes} K.~A.,  1990, \apj, 360, 561

\bibitem[\protect\citeauthoryear{{Longmore}, {Dixon}, {Skillen}, {Jameson} \&
  {Fernley}}{{Longmore} et~al.}{1990}]{long}
{Longmore} A.~J.,  {Dixon} R.,  {Skillen} I.,  {Jameson} R.~F.,    {Fernley}
  J.~A.,  1990, \mnras, 247, 684

\bibitem[\protect\citeauthoryear{{Mackey} \& {Gilmore}}{{Mackey} \&
  {Gilmore}}{2003}]{mac03}
{Mackey} A.~D.,  {Gilmore} G.~F.,  2003, \mnras, 345, 747

\bibitem[\protect\citeauthoryear{{Martin}, {Ibata}, {Bellazzini}, {Irwin},
  {Lewis} \& {Dehnen}}{{Martin} et~al.}{2004}]{mar04}
{Martin} N.~F.,  {Ibata} R.~A.,  {Bellazzini} M.,  {Irwin} M.~J.,  {Lewis}
  G.~F.,    {Dehnen} W.,  2004, \mnras, 348, 12

\bibitem[\protect\citeauthoryear{{Martin}, {Ibata}, {Conn}, {Lewis},
  {Bellazzini} \& {Irwin}}{{Martin} et~al.}{2005}]{mar05}
{Martin} N.~F.,  {Ibata} R.~A.,  {Conn} B.~C.,  {Lewis} G.~F.,  {Bellazzini}
  M.,    {Irwin} M.~J.,  2005, \mnras, 362, 906

\bibitem[\protect\citeauthoryear{{McLaughlin} \& {van der Marel}}{{McLaughlin}
  \& {van der Marel}}{2005}]{mcl05}
{McLaughlin} D.~E.,  {van der Marel} R.~P.,  2005, \apjs, 161, 304

\bibitem[\protect\citeauthoryear{{McWilliam} \& {Smecker-Hane}}{{McWilliam} \&
  {Smecker-Hane}}{2005}]{mcw05}
{McWilliam} A.,  {Smecker-Hane} T.~A.,  2005, \apjl, 622, L29

\bibitem[\protect\citeauthoryear{{Morrison}, {Mateo}, {Olszewski}, {Harding},
  {Dohm-Palmer}, {Freeman}, {Norris} \& {Morita}}{{Morrison}
  et~al.}{2000}]{mor00}
{Morrison} H.~L.,  {Mateo} M.,  {Olszewski} E.~W.,  {Harding} P.,
  {Dohm-Palmer} R.~C.,  {Freeman} K.~C.,  {Norris} J.~E.,    {Morita} M.,
  2000, \aj, 119, 2254

\bibitem[\protect\citeauthoryear{{Newberg}, {Yanny}, {Rockosi}, {Grebel},
  {Rix}, {Brinkmann}, {Csabai}, {Hennessy}, {Hindsley}, {Ibata}, {Ivezi{\'c}},
  {Lamb}, {Nash}, {Odenkirchen}, {Rave}, {Schneider}, {Smith}, {Stolte} \&
  {York}}{{Newberg} et~al.}{2002}]{new02}
{Newberg} H.~J.,  {Yanny} B.,  {Rockosi} C.,  {Grebel} E.~K.,  {Rix} H.,
  {Brinkmann} J.,  {Csabai} I.,  {Hennessy} G.,  {Hindsley} R.~B.,  {Ibata} R.,
   {Ivezi{\'c}} Z.,  {Lamb} D.,  {Nash} E.~T.,  {Odenkirchen} M.,  {Rave}
  H.~A.,  {Schneider} D.~P.,  {Smith} J.~A.,  {Stolte} A.,    {York} D.~G.,
  2002, \apj, 569, 245

\bibitem[\protect\citeauthoryear{{Pe{\~n}arrubia}, {Mart{\'{\i}}nez-Delgado},
  {Rix}, {G{\'o}mez-Flechoso}, {Munn}, {Newberg}, {Bell}, {Yanny}, {Zucker} \&
  {Grebel}}{{Pe{\~n}arrubia} et~al.}{2005}]{mon05}
{Pe{\~n}arrubia} J.,  {Mart{\'{\i}}nez-Delgado} D.,  {Rix} H.~W.,
  {G{\'o}mez-Flechoso} M.~A.,  {Munn} J.,  {Newberg} H.,  {Bell} E.~F.,
  {Yanny} B.,  {Zucker} D.,    {Grebel} E.~K.,  2005, \apj, 626, 128

\bibitem[\protect\citeauthoryear{{Peterson} \& {Cudworth}}{{Peterson} \&
  {Cudworth}}{1994}]{peterm22}
{Peterson} R.~C.,  {Cudworth} K.~M.,  1994, \apj, 420, 612

\bibitem[\protect\citeauthoryear{{Peterson}, {Rees} \& {Cudworth}}{{Peterson}
  et~al.}{1995}]{peterm4}
{Peterson} R.~C.,  {Rees} R.~F.,    {Cudworth} K.~M.,  1995, \apj, 443, 124

\bibitem[\protect\citeauthoryear{{Piatti} \& {Clari{\'a}}}{{Piatti} \&
  {Clari{\'a}}}{2008}]{pia08}
{Piatti} A.~E.,  {Clari{\'a}} J.~J.,  2008, \mnras, 390, L54

\bibitem[\protect\citeauthoryear{{Pickering}}{{Pickering}}{1908}]{pic08}
{Pickering} E.~C.,  1908, Annals of Harvard College Observatory, 60, 147

\bibitem[\protect\citeauthoryear{{Plummer}}{{Plummer}}{1911}]{plu11}
{Plummer} H.~C.,  1911, \mnras, 71, 460

\bibitem[\protect\citeauthoryear{{Pritzl}, {Smith}, {Catelan} \&
  {Sweigart}}{{Pritzl} et~al.}{2000}]{pritzl00}
{Pritzl} B.,  {Smith} H.~A.,  {Catelan} M.,    {Sweigart} A.~V.,  2000, \apjl,
  530, L41

\bibitem[\protect\citeauthoryear{{Robin}, {Reyl{\'e}}, {Derri{\`e}re} \&
  {Picaud}}{{Robin} et~al.}{2003}]{rob03}
{Robin} A.~C.,  {Reyl{\'e}} C.,  {Derri{\`e}re} S.,    {Picaud} S.,  2003,
  \aap, 409, 523

\bibitem[\protect\citeauthoryear{{Rutledge}, {Hesser} \& {Stetson}}{{Rutledge}
  et~al.}{1997}]{rutl97b}
{Rutledge} G.~A.,  {Hesser} J.~E.,    {Stetson} P.~B.,  1997, \pasp, 109, 907

\bibitem[\protect\citeauthoryear{{Salaris} \& {Weiss}}{{Salaris} \&
  {Weiss}}{2002}]{sal02}
{Salaris} M.,  {Weiss} A.,  2002, \aap, 388, 492

\bibitem[\protect\citeauthoryear{{Sandage}}{{Sandage}}{1993}]{sand93}
{Sandage} A.,  1993, \aj, 106, 687

\bibitem[\protect\citeauthoryear{{Sarajedini}}{{Sarajedini}}{1993}]{saraj93}
{Sarajedini} A.,  1993, \aj, 105, 2172

\bibitem[\protect\citeauthoryear{{Sharp}, {Saunders}, {Smith}, {Churilov},
  {Correll}, {Dawson}, {Farrel}, {Frost}, {Haynes}, {Heald}, {Lankshear},
  {Mayfield}, {Waller} \& {Whittard}}{{Sharp} et~al.}{2006}]{sharp06}
{Sharp} R.,  {Saunders} W.,  {Smith} G.,  {Churilov} V.,  {Correll} D.,
  {Dawson} J.,  {Farrel} T.,  {Frost} G.,  {Haynes} R.,  {Heald} R.,
  {Lankshear} A.,  {Mayfield} D.,  {Waller} L.,    {Whittard} D.,  2006, in
  Society of Photo-Optical Instrumentation Engineers (SPIE) Conference Series
  Vol.~6269 of Society of Photo-Optical Instrumentation Engineers (SPIE)
  Conference Series, {Performance of AAOmega: the AAT multi-purpose fiber-fed
  spectrograph}

\bibitem[\protect\citeauthoryear{{Shetrone}, {Venn}, {Tolstoy}, {Primas},
  {Hill} \& {Kaufer}}{{Shetrone} et~al.}{2003}]{shet03}
{Shetrone} M.,  {Venn} K.~A.,  {Tolstoy} E.,  {Primas} F.,  {Hill} V.,
  {Kaufer} A.,  2003, \aj, 125, 684

\bibitem[\protect\citeauthoryear{{Skrutskie}}{{Skrutskie}}{2006}]{twomass}
{Skrutskie} M.~F. e.~a.,  2006, \aj, 131, 1163

\bibitem[\protect\citeauthoryear{{Smith} \& {Drake}}{{Smith} \&
  {Drake}}{1990}]{drake90}
{Smith} G.,  {Drake} J.~J.,  1990, \aap, 231, 125

\bibitem[\protect\citeauthoryear{{Smith}}{{Smith}}{1984}]{smith84}
{Smith} H.~A.,  1984, \pasp, 96, 505

\bibitem[\protect\citeauthoryear{{Smith} \& {Perkins}}{{Smith} \&
  {Perkins}}{1982}]{perk}
{Smith} H.~A.,  {Perkins} G.~J.,  1982, \apj, 261, 576

\bibitem[\protect\citeauthoryear{{Storm}}{{Storm}}{2004}]{storm04}
{Storm} J.,  2004, \aap, 415, 987

\bibitem[\protect\citeauthoryear{{Tolstoy}, {Venn}, {Shetrone}, {Primas},
  {Hill}, {Kaufer} \& {Szeifert}}{{Tolstoy} et~al.}{2003}]{tol03}
{Tolstoy} E.,  {Venn} K.~A.,  {Shetrone} M.,  {Primas} F.,  {Hill} V.,
  {Kaufer} A.,    {Szeifert} T.,  2003, \aj, 125, 707

\bibitem[\protect\citeauthoryear{{Tonry} \& {Davis}}{{Tonry} \&
  {Davis}}{1979}]{ton79}
{Tonry} J.,  {Davis} M.,  1979, \aj, 84, 1511

\bibitem[\protect\citeauthoryear{{Trager}, {Djorgovski} \& {King}}{{Trager}
  et~al.}{1993}]{trag93}
{Trager} S.~C.,  {Djorgovski} S.,    {King} I.~R.,  1993, in {S.~G.~Djorgovski
  \& G.~Meylan} ed., Structure and Dynamics of Globular Clusters Vol.~50 of
  Astronomical Society of the Pacific Conference Series, {Structural Parameters
  of Galactic Globular Clusters}.
pp 347--+

\bibitem[\protect\citeauthoryear{{van den Bergh}}{{van den
  Bergh}}{1993}]{van93}
{van den Bergh} S.,  1993, \aj, 105, 971

\bibitem[\protect\citeauthoryear{{Venn}, {Irwin}, {Shetrone}, {Tout}, {Hill} \&
  {Tolstoy}}{{Venn} et~al.}{2004}]{venn04}
{Venn} K.~A.,  {Irwin} M.,  {Shetrone} M.~D.,  {Tout} C.~A.,  {Hill} V.,
  {Tolstoy} E.,  2004, \aj, 128, 1177

\bibitem[\protect\citeauthoryear{{Vivas}}{{Vivas}}{2001}]{viv01}
{Vivas} A.~K. e.~a.,  2001, \apjl, 554, L33

\bibitem[\protect\citeauthoryear{{Walker} \& {Nemec}}{{Walker} \&
  {Nemec}}{1996}]{walker96}
{Walker} A.~R.,  {Nemec} J.~M.,  1996, \aj, 112, 2026

\bibitem[\protect\citeauthoryear{{Warren} \& {Cole}}{{Warren} \&
  {Cole}}{2009}]{war09}
{Warren} S.~R.,  {Cole} A.~A.,  2009, \mnras, 393, 272

\bibitem[\protect\citeauthoryear{{Yanny}}{{Yanny}}{2000}]{yan00}
{Yanny} B. e.~a.,  2000, \apj, 540, 825

\bibitem[\protect\citeauthoryear{{Younger}, {Besla}, {Cox}, {Hernquist},
  {Robertson} \& {Willman}}{{Younger} et~al.}{2008}]{you08}
{Younger} J.~D.,  {Besla} G.,  {Cox} T.~J.,  {Hernquist} L.,  {Robertson} B.,
   {Willman} B.,  2008, \apjl, 676, L21

\bibitem[\protect\citeauthoryear{{Zinn} \& {West}}{{Zinn} \&
  {West}}{1984}]{zinn}
{Zinn} R.,  {West} M.~J.,  1984, \apjs, 55, 45

\end{thebibliography}

\clearpage

\begin{table}
\begin{center}
\caption{Log of Observations\label{tab-log}}
\begin{tabular}{lcccccccr}
\hline\hline
Target & $\alpha$ (J2000)$^a$ & $\delta$ (J2000)$^a$ & UT start & Airmass & Seeing ($^{\prime\prime}$) & t$_{exp}$ (s)
& [Fe/H] (dex)$^b$ & V$_r$ (km/s) \\
\hline
M68       & 12 39 28 & $-$25 15 39 & 11:43:55 & 1.06 & 1.4 & 2$\times$360  & $-$1.99 $\pm$0.06 & $-$96.4 $\pm$3.9$^c$ \\
M4        & 16 23 34 & $-$26 32 01 & 18:16:28 & 1.72 & 1.4 & 2$\times$180  & $-$1.19 $\pm$0.03 & 70.9 $\pm$0.6$^d$ \\
M22       & 18 36 25 & $-$23 54 16 & 19:14:59 & 1.32 & 1.4 & 2$\times$180  & $-$1.48 $\pm$0.03$^{\dagger}$ & $-$148.8 $\pm$0.8$^e$\\
IC~4499 1 & 15 00 21 & $-$82 12 46 & 13:09:58 & 1.60 & 1.4 & 2$\times$1800 &  & \\
IC~4499 2 & 15 00 22 & $-$82 12 52 & 15:34:27 & 1.67 & 1.4 & 2$\times$1800 &  & \\
\hline
\end{tabular}
\end{center}
$^a$Centre of AAOmega/2dF field; $^b$\citet{cg97};
  $^c$\citet{geisler}; $^d$\citet{peterm4}; $^e$\citet{peterm22}.
$^{\dagger}$A significant range is present \citep{dac09}.
\end{table}

\begin{table}
\begin{center}
\caption{Summary of results.\label{tab-res}}
\begin{tabular}{lcccrrrr}
\hline\hline
Cluster & N$_{\star}$ & W$^{\prime}$ (\AA) & K$_{HB}$ (mag) & [Fe/H] & $\Delta$[Fe/H] & 
V$_r$ (km/s) & $\Delta$V$_r$ (km/s) \\
\hline
M68     & 51 & 2.59 $\pm$0.35 & 14.4$^a$ & $-$1.88 $\pm$0.13 &    0.11 $\pm$0.14 
&  $-$98.6 $\pm$1.5 & $-$4.2 $\pm$4.2 \\
M4      & 70 & 4.90 $\pm$0.34 & 11.13    & $-$1.12 $\pm$0.14 &    0.07 $\pm$0.14 
&     65.7 $\pm$0.9 &    5.2 $\pm$1.1 \\
M22     & 81 & 3.61 $\pm$0.46 & 12.21    & $-$1.55 $\pm$0.17 & $-$0.07 $\pm$0.17 
& $-$150.5 $\pm$1.3 & $-$1.7 $\pm$1.5 \\
IC~4499 & 43 & 3.70 $\pm$0.29 & 15.97    & $-$1.52 $\pm$0.12 &                   
&     31.5 $\pm$0.4 &                 \\
\hline
\end{tabular}
\end{center}
$^a$\citet{dallora}. $\Delta$ Difference, measured$-$literature value.
\end{table}

\clearpage

\begin{table}
\begin{center}
\caption{IC~4499 Members.\label{tab-stars}}
\begin{tabular}{llllcc}
\hline\hline
ID  & $\alpha$(J2000)$^a$ & $\delta$(J2000)$^a$ & V$_r$ (km/s) & $\Sigma$W (\AA ) & K$_S$$^a$ (mag) \\
\hline
4976&14:59:33.41&-82:09:17.80&31.40$\pm$1.49&5.51$\pm$0.36&12.55\\
4983&15:01:05.98&-82:12:36.93&31.21$\pm$1.49&5.00$\pm$0.52&12.65\\
5034&14:58:38.62&-82:10:30.86&32.42$\pm$1.46&4.27$\pm$0.66&14.22\\
5420&15:00:42.16&-82:08:32.40&32.88$\pm$1.93&4.73$\pm$0.90&13.98\\
5437&15:01:49.80&-82:13:39.74&31.16$\pm$1.48&5.06$\pm$0.94&14.47\\
5447&14:59:01.37&-82:10:50.80&28.91$\pm$1.96&5.07$\pm$0.70&13.57\\
5478&14:58:54.40&-82:16:34.04&33.44$\pm$1.46&4.58$\pm$0.56&13.79\\
5488&15:02:14.17&-82:15:35.02&30.88$\pm$1.49&4.95$\pm$1.33&14.70\\
5595&14:59:46.75&-82:16:15.61&30.47$\pm$1.49&5.64$\pm$0.46&12.31\\
5644&14:59:40.00&-82:12:22.84&27.48$\pm$1.48&4.77$\pm$0.68&13.89\\
5649&14:58:19.16&-82:08:30.86&30.89$\pm$1.66&6.03$\pm$0.45&11.61\\
5914&14:58:56.49&-82:13:14.09&32.28$\pm$1.46&3.80$\pm$0.81&14.22\\
5916&14:58:54.78&-82:03:44.48&31.29$\pm$1.67&5.72$\pm$0.49&12.16\\
5971&15:00:38.30&-82:11:12.03&30.18$\pm$1.40&6.60$\pm$0.49&10.77\\
6150&14:59:50.74&-82:13:10.60&30.86$\pm$1.49&4.32$\pm$0.61&13.81\\
6210&14:59:18.66&-82:12:40.50&30.62$\pm$1.93&5.45$\pm$0.46&12.64\\
6266&15:00:05.66&-82:15:51.72&34.06$\pm$1.65&4.24$\pm$1.01&14.78\\
6302&15:00:10.58&-82:11:17.03&31.88$\pm$1.50&5.70$\pm$0.40&12.22\\
6370&14:57:18.38&-82:10:41.52&31.84$\pm$1.48&4.43$\pm$0.84&14.51\\
6389&15:01:25.51&-82:20:14.43&30.99$\pm$1.49&5.61$\pm$0.27&11.99\\
6450&15:01:55.59&-82:10:45.69&34.28$\pm$1.49&5.10$\pm$0.47&12.59\\
6478&14:59:45.22&-82:14:47.23&29.58$\pm$1.47&4.88$\pm$0.57&13.76\\
6688&15:04:56.01&-82:20:08.58&33.09$\pm$1.63&5.19$\pm$1.45&14.79\\
6689&15:01:21.54&-82:13:45.49&31.71$\pm$1.30&4.56$\pm$0.91&14.41\\
6693&15:01:31.47&-82:12:24.68&29.63$\pm$1.44&3.93$\pm$0.91&14.49\\
6698&15:00:17.12&-82:16:35.09&29.65$\pm$1.48&5.00$\pm$0.37&13.10\\
6703&15:00:34.26&-82:14:45.50&38.08$\pm$1.49&5.52$\pm$0.40&11.35\\
6710&14:59:47.84&-82:14:13.06&36.36$\pm$1.39&4.02$\pm$1.08&14.75\\
6718&15:00:06.81&-82:11:52.33&26.83$\pm$1.48&4.54$\pm$0.62&13.93\\
6732&15:01:05.76&-82:12:57.37&28.30$\pm$1.50&4.82$\pm$0.69&13.79\\
6847&15:00:51.51&-82:12:50.99&34.76$\pm$1.58&4.07$\pm$1.39&14.97\\
6850&15:02:54.39&-82:11:43.51&31.56$\pm$1.93&5.42$\pm$0.50&12.93\\
7024&14:58:40.81&-82:10:36.05&31.65$\pm$1.50&4.01$\pm$1.14&14.09\\
7088&14:59:55.26&-82:13:13.16&29.35$\pm$1.46&4.68$\pm$0.82&14.65\\
7089&15:00:58.89&-82:14:09.08&30.09$\pm$1.94&4.58$\pm$0.52&13.64\\
7126&14:58:59.03&-82:08:33.25&29.97$\pm$1.92&4.84$\pm$0.60&13.49\\
7162&15:00:47.33&-82:12:51.62&35.64$\pm$1.51&4.38$\pm$0.93&14.06\\
7290&15:00:39.11&-82:13:17.03&25.95$\pm$1.48&4.83$\pm$0.73&14.11\\
7508&14:59:42.17&-82:10:05.79&31.96$\pm$2.00&4.72$\pm$2.02&14.16\\
7529&15:00:17.01&-82:11:11.46&31.52$\pm$1.71&5.79$\pm$0.41&11.94\\
7558&14:57:40.75&-82:10:33.69&30.41$\pm$1.49&5.77$\pm$1.03&11.58\\
7575&14:59:40.44&-82:16:04.08&31.52$\pm$1.48&5.98$\pm$0.33&11.78\\
7910&14:59:47.29&-82:11:09.15&33.05$\pm$1.50&4.68$\pm$0.56&14.06\\
\hline
\end{tabular}
\end{center}
$^a$From 2MASS point source catalogue.
\end{table}

\end{document}